\documentclass[12pt]{article}
\usepackage{amstext}
\usepackage{amssymb}
\usepackage{amsmath}

\usepackage{cite}
\usepackage{color}

\usepackage{graphicx}
\usepackage{float}

\usepackage[cp1251]{inputenc}
\usepackage[english]{babel}

\usepackage{version}

\renewcommand{\theequation}{\arabic{section}.\arabic{equation}}%

\begin{document}

\large
\makeatletter
\renewcommand{\@oddhead}{\hfil\thepage\hfil}
\makeatother

\thispagestyle{empty}
\begin{center}
\large{\bf Thin Film Motion of an Ideal Fluid \\ on the
Rotating Cylinder Surface}

\end{center}

\begin{center}
\large{\bf M. Yu. Zhukov$^{a}$ and A. M. Morad$^{b,}$$^{*}$}

\bigskip

   \small{\it Faculty of Mathematics, Mechanics and Computer Sciences \\ Southern Federal University \\ Rostov-on-Don,  344090 Russia}

\small{\it $^{*}$\ Department of Mathematics, Faculty of Science, Menoufiya University, 32511 Egypt}

   {\small \textit{E-mail:} $^{a}$\ zhuk@math.rsu.ru,$^{b}$\ am.morad@menofia.edu.eg  }

\end{center}

\vspace{\baselineskip}
{\small
 {\bf Abstract}

\medskip
The shallow water equations describing the motion of thin liquid film on the rotating cylinder surface are obtained.
These equations are the analog of the modified Boussinesq equations for shallow water and the Korteweg-de Vries equation. It is clear that for rotating cylinder the centrifugal force plays the role of the gravity. For construction the shallow water equations (amplitude equations) usual depth-averaged and multi-scale asymptotic expansion methods are used. Preliminary analysis shows that a thin film of an ideal incompressible fluid precesses around the axis of the cylinder with velocity which differs from the angular velocity of rotating cylinder. For the mathematical model of the liquid film motion the analytical solutions are obtained by the Tanh-Function method. To illustrate the integrability of the equations the Painlev\'{e} analysis is used. The truncated expansion method  and symbolic computation allows to present an auto-B\"{a}cklund transformation. The results of analysis show that the exact solutions of the model correspond to the solitary waves of different types.
}

\section*{Introduction}

\medskip
Many scientific and industrial problems connect to the flow of thin liquid films. Thin film technology is used extensively in many applications including microelectronics, optics, magnetic, hard and corrosion resistant coatings, micro-mechanics, biotechnology, medicine, laser, etc. At larger scales the ascent of buoyant magma below solid rocks and the spreading of lava on volcanoes are further examples of geological problems \cite{AbourabiaHassanMorad}. Progress in these areas depends upon the comprehension of fluid flow mechanisms.
As a rule the behavior of thin liquid films of an incompressible ideal fluid can be described by the shallow water equations. Classic shallow water equations are obtained by the depth-averaging of the Euler equations for an incompressible fluid under the assumption of potential flow (see, for instance \cite{KortewegdeVries,Newell,Rayleigh,LyapidevskyTeshukov,Whithem,
OvsyannikovMakarenkoNalimov}).  In this case the gravity  plays the significant role.

The main objective of this paper is to construct the shallow water equations for thin liquid layer (film) coated the surface of the infinitely long cylinder rotating with a constant angular velocity. In this case the role of gravity plays the centrifugal force. The usual depth-averaged technique and the multiscale asymptotic expansions method allows to obtain the analogues of the Boussinesq shallow water model  and the Korteweg-de Vries equation. The main difference between classic equation and obtained model is the presence of the liquid layer curvature since the liquid motion occurs on the surface of a cylinder of finite radius. Naturally, with the tendency of a radius of cylinder to infinity the presented model pass into the classical model. Other difference from the classical model is the presence of the vortex flow with constant vorticity (so-called the vortex shallow water equations, see for instance \cite{LyapidevskyTeshukov,OvsyannikovMakarenkoNalimov}).

Other objective of this paper is the investigation of analytic solutions for obtained equations.  The study of exact solutions of the nonlinear PDEs has become one of the most important topics in mathematical physics. In the past decades, various powerful methods like the Inverse scattering method, variable separation approach and Homogeneous balance method were used. But, in recent years, much research works has been concentrated on the Cole-Hopf transformation method, the Jacobi elliptic method, Adomian method, and the various extensions of the Tanh-function method \cite{AbourabiaHorbaty,ZayedAbourabiaGepreelHorbaty,ZayedRahman,AbourabiaDanafMorad,
Wazwaz,Soliman,Khuri}.

The paper is organized as follows. In Sec.~\ref{zhmor:sec:2} we  introduce the basic equations governing the thin liquid film flow on the surface of a rotating cylinder. The corresponding two-dimensional free boundary problem is presented in Sec.~\ref{zhmor:sec:3}. In Sec.~\ref{zhmor:sec:4} we apply the perturbation technique and the suitable transformations to construct a system of hyperbolic equations. In Sec.~\ref{zhmor:sec:5}  we derive the KdV equation on the basis of the Boussinesq model using ordinary technic of the amplitude equation constructing (see, for instance \cite{Gibbon}). In Sec.~\ref{zhmor:sec:6} we present explicit Painlev\'{e} test for the model equation. In Sec.~\ref{zhmor:sec:7} we solve the model equation analytically by using two different methods and discuss the fundamental properties of the model.

\bigskip

\setcounter{equation}{0}

\section{The Basic Equations}\label{zhmor:sec:2}

\begin{figure}[H]
\centering  \includegraphics[scale=1]{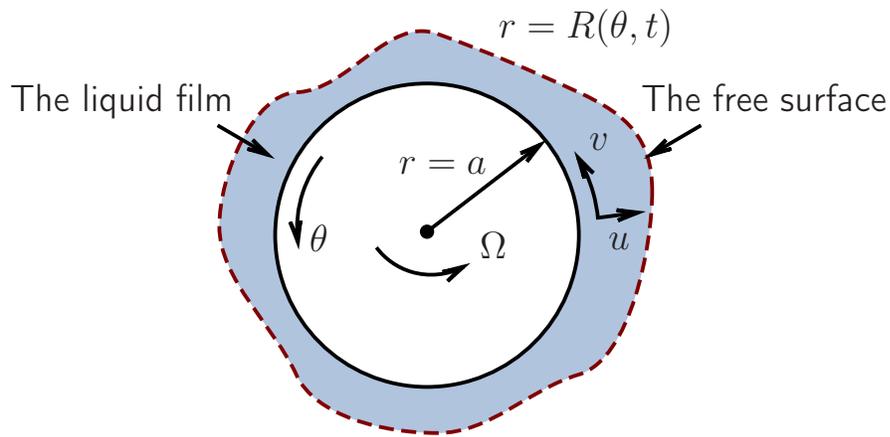}\\
  \caption{The sketch of liquid film which coat the rotating cylinder}\label{cylinder}
\end{figure}

We assume that a thin layer of the ideal incompressible fluid with the free boundary coats the surface of the infinitely long cylinder rotating with a constant angular velocity (see Fig.\,\ref{cylinder}). To describe the behavior of this layer the Euler equation and the continuity equation rewritten in cylindrical coordinates $(r,\theta)$ are used
\begin{equation}
u_t+uu_r + \frac{1}{r}vu_\theta - \frac{v^2}{r}=-p_r,
\label{2.1}
\end{equation}
\begin{equation}
v_t+uv_r + \frac{1}{r}vv_\theta + \frac{uv}{r}=-\frac{1}{r}p_\theta,
\label{2.2}
\end{equation}
\begin{equation}
(ru)_r+v_\theta = 0, \label{2.3}
\end{equation}
\begin{equation}
D=\{ a < r < R(\theta,t),\quad 0 < \theta < 2\pi\}.
\end{equation}
Here $D$  is the
region filled by the liquid, $u$ is the radial velocity, $v$ is the angular velocity, $p$ is the
pressure, $a$ is the radius of cylinder.

The equation of the free liquid surface is
\begin{equation}
r=R(\theta,t),
\label{2.4}
\end{equation}
where $R(\theta,t)$ is an
unknown function that defines the free liquid surface.

The impermeability condition, the kinematic and
dynamic conditions on the free boundary have the following form
\begin{equation}
u=0, \quad r=a,
\label{2.5}
\end{equation}
\begin{equation}
R_t+\frac{1}{r}v R_\theta-u=0, \quad r=R(\theta,t),
\label{2.6}
\end{equation}
\begin{equation}
p=\Pi(\theta,t),\quad r=R(\theta,t),
\label{2.7}
\end{equation}
where $\Pi(\theta,t)$ is the pressure on the free boundary.

Naturally, no-slip conditions at the boundary of the
liquid-cylinder are absent because of an ideal fluid.

The dimensional and dimensionless variables are
connected by formulae (dimensional variables marked by asterisk)
\begin{equation}
t=\Omega_\ast t_\ast, \quad
   r_\ast=R_\ast r, \quad
   R_\ast^{in}=R_\ast a, \quad
   \Omega_\ast^0=\Omega \Omega_\ast,   \nonumber
\end{equation}
\begin{equation}
(u^\ast,v^\ast)=\Omega_\ast R_\ast (u,v),\quad
   p_\ast=\rho_\ast \Omega_\ast^2 R_\ast^2p
\label{2.8}
\end{equation}
Here $\Omega_\ast$, $\Omega_\ast^0$, $R_\ast$,
$R_\ast^{in}$, $\rho_\ast$, $\nu_\ast$ are
the characteristic angular velocity, the angular velocity
of the fluid, the characteristic radius, the inner radius of cylinder, the liquid density, and the kinematic viscosity, respectively.

\setcounter{equation}{0}

\section{Fluid Flow with a Constant Vortex}\label{zhmor:sec:3}

In this section we construct so called the vortex shallow water equations (see, for instance, \cite{LyapidevskyTeshukov,OvsyannikovMakarenkoNalimov}). We assume that vortex of the fluid flow $\omega$ equals to $2\Omega$
\begin{equation}
\omega=\frac{1}{r}(rv)_r-\frac{1}{r}u_\theta, \quad \omega=2\Omega={\rm
const}.
\label{3.1}
\end{equation}
Obviously, the case of the potential fluid flow corresponds to $\Omega=0$.

We introduce the stream function
\begin{equation}
u=-r \psi_\theta, \quad v=\psi_r.
\label{3.2}
\end{equation}

To determine the stream function $\psi$ we have the following
underdetermined problem
\begin{equation}
\Delta \psi \equiv \frac{1}{r}\frac{\partial }{\partial r} r
\frac{\partial\psi}{\partial r}
+\frac{1}{r^2}\psi_{\theta\theta}=2\Omega,
\label{3.3}
\end{equation}
\begin{equation}
\psi=0, \quad r=a
\label{3.4}
\end{equation}

It is easy to show that the solution of the problem (\ref{3.3}), (\ref{3.4}) can be written as (see, for instance, \cite{Newell} and Appendix 1)
\begin{equation}
\displaystyle\psi(r,\theta,t)=\frac{1}{2}\Omega(r^2-a^2)+
\sum\limits_{j=0}^{\infty}\,(-1)^{j} \frac{\left(\ln
\left(\displaystyle\frac{r}{a}\right) \right)^{2j+1}}{(2j+1)!}
\frac{\partial^{2j}}{\partial\theta^{2j}}F(\theta,t),
\label{3.5}
\end{equation}
where $F(\theta,t)$ is an arbitrary function.

We denote the stream function at the free boundary as
\begin{equation}
\Psi(\theta,t)=\psi(R(\theta,t),\theta,t).
 \label{3.6}
\end{equation}
Then the kinematic condition at the free boundary (\ref{2.6}) can be written in the following form
\begin{equation}
RR_t+\Psi_\theta=0.
 \label{3.7}
\end{equation}

Using the dynamic condition at the free boundary (\ref{2.7}) and
(\ref{2.2}), (\ref{2.3}), we get
\begin{eqnarray}
  &\displaystyle  (rv)_t+u(rv)_r+\frac{v}{r}(rv)_\theta+ R_\theta \left( u_t+uu_r +
  \frac{v}{r}u_\theta - \frac{v^2}{r} \right)=-\Pi_\theta,  \label{3.8} \\
  &\displaystyle r=R(\theta,t). \nonumber
\end{eqnarray}

The system of equations (\ref{3.7}), (\ref{3.8}) with  (\ref{3.2}), (\ref{3.5}), (\ref{3.6}) is a closed system of equations for determining the
functions $R(\theta,t)$, $F(\theta,t)$.

\setcounter{equation}{0}

\section{Long-wave Approximation}\label{zhmor:sec:4}

Using the ordinary technique of constructing multi-scale asymptotic expansions we introduce a fast variable $\eta$, $\tau$ and change scale of the variables
\begin{equation}
\tau=\varepsilon t, \quad \eta=\varepsilon \theta,
\quad u=\varepsilon U,
\label{4.1}
\end{equation}
where  $\varepsilon$ is the small
parameter related to the thickness of the liquid layer.

In this case, equations (\ref{3.7}),(\ref{3.8}),(\ref{3.2}),(\ref{3.5}),(\ref{3.6}) can be rewritten as
\begin{equation}\label{4.2}
  RR_\tau+\Psi_\eta=0,
\end{equation}
\begin{eqnarray}
&&(rv)_\tau+U(rv)_r+\frac{v}{r}(rv)_\eta-R_\eta~\frac{v^2}{r}+{}\nonumber\\
&&{}+\varepsilon^2 R_\eta
\left(
U_\tau+UU_r + \frac{v}{r}U_\eta
\right)=-\Pi_\eta, \quad  r=R(\eta,\tau),
\label{4.3}
\end{eqnarray}
\begin{equation}\label{4.4}
 \Psi(\eta,\tau)=\frac{1}{2}\Omega(R^2-a^2)+
\sum\limits_{j=0}^{\infty}\,(-1)^{j}~\varepsilon^{2j}~
\frac{Z^{2j+1}}{(2j+1)!}~
\frac{\partial^{2j}}{\partial \eta^{2j}}
\,F(\eta,\tau),
\end{equation}
\begin{equation}\label{4.5}
-rU=\psi_\eta=
\sum\limits_{j=0}^{\infty}\,(-1)^{j}~\varepsilon^{2j}~
\frac{z^{2j+1}}{(2j+1)!}
~\frac{\partial^{2j+1}}{\partial \eta^{2j+1}}
\,F(\eta,\tau),
\end{equation}
\begin{equation}\label{4.6}
rv=r\psi_r=\Omega r^2+
\sum\limits_{j=0}^{\infty}\,(-1)^{j}~\varepsilon^{2j}~
\frac{z^{2j}}{(2j)!}
~\frac{\partial^{2j}}{\partial \eta^{2j}}
\,F(\eta,\tau),
\end{equation}
\begin{equation}\label{4.7}
  z=\ln\left(\frac{r}{a}\right), \quad Z=\ln\left(\frac{R}{a}\right).
\end{equation}

For convenience we rewrite
(\ref{4.4})--(\ref{4.7}) omitting
terms of order more than $O(\varepsilon^2)$
\begin{equation}\label{4.8}
 \Psi(\eta,\tau)=
\frac{1}{2}\Omega(R^2-a^2)+
ZF-\varepsilon^2\frac{Z^3}{3!}F_{\eta\eta}+O(\varepsilon^4),
\end{equation}
\begin{equation}\label{4.9}
 -rU=\psi_\eta=
zF_\eta-\varepsilon^2\frac{z^3}{3!}F_{\eta\eta\eta}+O(\varepsilon^4),
\end{equation}
\begin{equation}\label{4.10}
 rv=r\psi_r=\Omega r^2+F-\varepsilon^2\frac{z^2}{2!}F_{\eta\eta}+O(\varepsilon^4),
\end{equation}
\begin{equation}\label{4.11}
 (rv)_\tau=F_\tau-\varepsilon^2\frac{z^2}{2!}F_{\eta\eta\tau}+O(\varepsilon^4),
\end{equation}
\begin{equation}\label{4.12}
(rv)_r=2\Omega r-\varepsilon^2\frac{z}{r}F_{\eta\eta}+O(\varepsilon^4),
\end{equation}
\begin{equation}\label{4.13}
 (rv)_\eta=F_\eta-\varepsilon^2\frac{z^2}{2!}F_{\eta\eta\eta}+O(\varepsilon^4),
\end{equation}
\begin{equation}\label{4.14}
 U_\tau=-~\frac{z}{r}F_{\eta\tau}+O(\varepsilon^2),
\end{equation}
\begin{equation}\label{4.15}
  U_r=-\left(\frac{1}{r^2}-\frac{z}{r^2}\right)F_{\eta}+O(\varepsilon^2),
\end{equation}
\begin{equation}\label{4.16}
 U_\eta=-~\frac{z}{r}F_{\eta\eta}+O(\varepsilon^2),
\end{equation}
\begin{equation}\label{4.17}
  \Psi_\eta(\eta,\tau)=
\Omega RR_\eta+
\left(
ZF-\varepsilon^2\frac{Z^3}{3!}F_{\eta\eta}
\right)_\eta+
O(\varepsilon^4).
\end{equation}

Substituting  (\ref{4.8})--(\ref{4.17}) into (\ref{4.3}) we obtain
\begin{eqnarray*}
&&F_\tau-\varepsilon^2\frac{Z^2}{2!}F_{\eta\eta\tau}- \frac{1}{R} \left(
ZF_\eta-\varepsilon^2\frac{Z^3}{3!}F_{\eta\eta\eta} \right) \left(
2\Omega R-\varepsilon^2\frac{Z}{R}F_{\eta\eta}
\right)+\nonumber\\
&&+\frac{1}{R^2}
\left(
\Omega R^2+F-\varepsilon^2\frac{Z^2}{2!}F_{\eta\eta}
\right)
\left(
F_\eta-\varepsilon^2\frac{Z^2}{2!}F_{\eta\eta\eta}
\right)-\nonumber\\
&&-\frac{R_\eta}{R^3}
\left(
\Omega R^2+F-\varepsilon^2\frac{Z^2}{2!}F_{\eta\eta}
\right)^2+\nonumber\\
&&+\varepsilon^2 R_\eta
\left\{
-~\frac{Z}{R}F_{\eta\tau}+
\frac{Z}{R}F_{\eta}
\left(\frac{1}{R^2}-\frac{Z}{R^2}\right)F_{\eta}-
\frac{1}{R^2}
\left(
\Omega R^2+F
\right)
\frac{Z}{R}F_{\eta\eta}
\right\}
=-\Pi_\eta.
\end{eqnarray*}

Retaining terms of order $O(\varepsilon^2)$ we have
\begin{eqnarray*}
&&F_\tau-\varepsilon^2\frac{Z^2}{2!}F_{\eta\eta\tau}- 2\Omega ZF_\eta +
\varepsilon^2 \frac{Z^2}{R^2}F_\eta F_{\eta\eta} +
2\Omega \varepsilon^2 \frac{Z^3}{3!}F_{\eta\eta\eta} +\nonumber\\
&&+
\left(
\Omega+\frac{F}{R^2}
\right)F_\eta
-\varepsilon^2
\left(
\Omega+\frac{F}{R^2}
\right)
\frac{Z^2}{2!}F_{\eta\eta\eta}
-\frac{\varepsilon^2}{R^2}
\frac{Z^2}{2!}F_{\eta\eta}
F_\eta -\nonumber\\
&&-\frac{R_\eta}{R^3}
\left(\Omega R^2+F\right)^2
+\varepsilon^2\frac{R_\eta}{R^3}
\left(
\Omega R^2+F
\right)
Z^2 F_{\eta\eta} +\nonumber\\
&&+\varepsilon^2 R_\eta
\left\{
-~\frac{Z}{R}F_{\eta\tau}+
\frac{Z}{R}F_{\eta}
\left(\frac{1}{R^2}-\frac{Z}{R^2}\right)F_{\eta}-
\frac{1}{R^2}
\left(
\Omega R^2+F
\right)
\frac{Z}{R}F_{\eta\eta}
\right\}=
-\Pi_\eta.
\end{eqnarray*}

Finally, substituting (\ref{4.8})--(\ref{4.17}) into (\ref{4.2}) and regrouping terms we get Boussinesq equations for describing the
vortex shallow water
\begin{equation}
 RR_\tau+\Omega RR_\eta+ \left(ZF\right)_\eta=
\varepsilon^2\left( \frac{Z^3}{3!}F_{\eta\eta} \right)_\eta, \quad
Z=\ln\left(\frac{R}{a}\right),
\label{4.18}
\end{equation}
\begin{eqnarray}
&& F_\tau+
\left(
-2\Omega ZF-\frac{1}{2}\Omega^2R^2+\Omega F + \frac{1}{2}~\frac{F^2}{R^2}
\right)_\eta=
\label{4.19}\\
&&=\varepsilon^2\left(
\frac{Z^2}{2}F_{\eta\tau}-\frac{Z^2F_\eta^2}{2R^2}
+\frac{Z^2}{2}F_{\eta\eta}\left(\Omega+\frac{F}{R^2}\right)- 2\Omega\frac{Z^3}{3!}F_{\eta\eta}
\right)_\eta-\Pi_\eta.
\nonumber
\end{eqnarray}

If we neglect terms of order $O(\varepsilon^2)$ (i.e., the dispersion
terms) then we obtain system of hyperbolic conserve laws
\begin{equation}
\left(\frac{1}{2}R^2\right)_\tau+
\left(\frac{1}{2}\Omega R^2+ZF\right)_\eta=0,
\label{4.20}
\end{equation}
\begin{equation}
F_\tau+
\left(
-2\Omega ZF-\frac{1}{2}\Omega^2R^2+\Omega F + \frac{1}{2}~\frac{F^2}{R^2}
\right)_\eta=0.
\label{4.21}
\end{equation}

Note that the investigation of this system has the independent interest. In particular, we can construct a self-similar solutions, rarefaction and  shock waves, and solve the Riemann problem
\cite{Rozhdestvenskii,Whithem}

\setcounter{equation}{0}

\section{The Korteweg-de Vries Equation}\label{zhmor:sec:5}

Using the ordinary technique for constructing of amplitude equation (see, for instance \cite{Gibbon}) we can
simplify the Boussinesq equations (\ref{4.18}),
(\ref{4.19}). We consider solution of these equations in the neighborhood of some characteristics of the linearized hyperbolic equations. Note that it is possible to construct models of different complexity using additional assumptions about the
magnitude of the free surface perturbation.

\begin{figure}[H]
\centering  \includegraphics[scale=1.0]{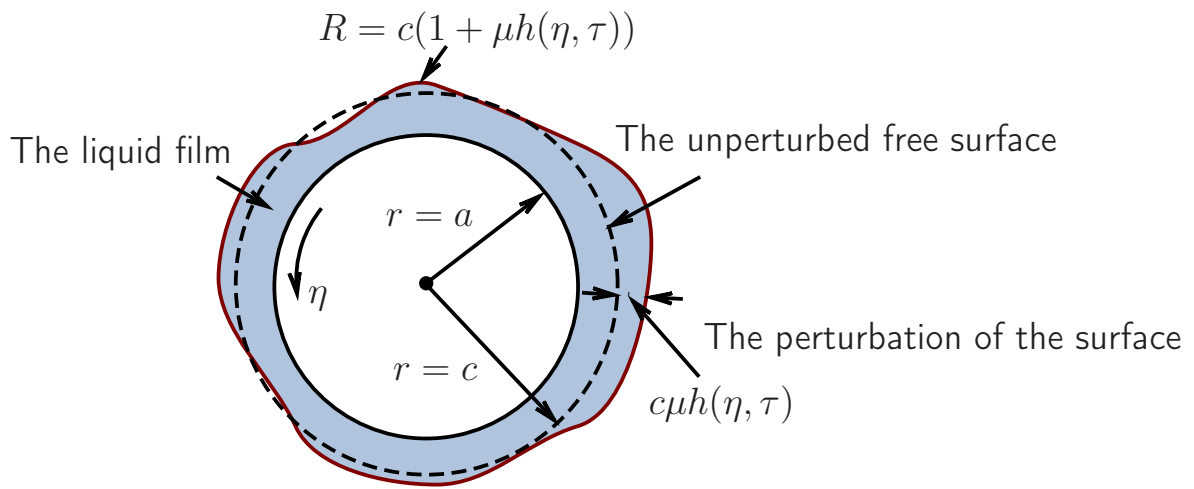}\\
  \caption{The deviation from the unperturbed free surface $r=c$}\label{cylinder2}
\end{figure}

From our point of view we consider the most interesting case when the unperturbed free surface is a cylinder of radius $c>a$ and the deviation of the free surface is sufficiently small (see fig.\,\ref{cylinder2})
\begin{equation}
R(\eta,\tau)=c(1+\mu h(\eta,\tau)).
\label{5.1}
\end{equation}
Here  $\mu h(\eta,\tau)$ is the function characterizing the deviation from the surface $r=c$, $\mu$ is the parameter characterizing the magnitude of the deviation.

We also assume that the function $F=O(\mu)$
\begin{equation}
F(\eta,\tau)=\mu w(\eta,\tau). \label{5.2}
\end{equation}

In general case $\varepsilon$ and $\mu$ are independent parameters. However, for simplicity we can link these parameters with help relation
\begin{equation}
\mu=\varepsilon^2.
\label{5.3}
\end{equation}
It means that magnitude of the free liquid surface perturbation is considerably less than the averaging thickness of the liquid.

We express the function $Z$ in the following form
\begin{equation}
Z=\ln\left(\frac{R}{a}\right)=Z_0+\mu h +O(\mu^2),
\quad Z_0=\ln \frac{c}{a}.
\label{5.4}
\end{equation}
Substituting (\ref{5.1})--(\ref{5.4}) into the Boussinesq equations
(\ref{4.18}), (\ref{4.19}) and keeping only terms of order $O(\varepsilon^2)$ and $O(\mu)$ we get

\begin{equation}
c^2(1+\mu h)h_\tau + \Omega c^2(1+\mu h)h_\eta+
\left((Z_0+\mu h)w\right)_\eta= \frac{1}{6}\varepsilon^2 Z_0^3
w_{\eta\eta\eta},
\label{5.5}
\end{equation}

\begin{eqnarray}
&& w_\tau+
(-2\Omega (Z_0+\mu h) w)_\eta -\Omega^2 c^2 (1+ \mu h) h_\eta +\Omega w_\eta +
\frac{\mu}{c^2} w w_\eta
=\nonumber\\
&&=\varepsilon^2\left(
\frac{1}{2}w_{\eta\eta\tau}
+\frac{1}{6}\Omega w_{\eta\eta\eta}
\right)-\Pi_\eta.
\label{5.6}
\end{eqnarray}

Omitting terms of order $O(\varepsilon^2)$ and $O(\mu)$ we obtain a
system of the first-order linear PDEs (not
necessarily hyperbolic!)
\begin{equation}
h_\tau + \Omega h_\eta+
\frac{Z_0}{c^2} w_\eta=0,
\label{5.7}
\end{equation}
\begin{equation}
w_\tau
-\Omega^2 c^2 h_\eta +(\Omega-2\Omega Z_0)w_\eta =0.
\label{5.8}
\end{equation}

To determine the characteristic directions $\lambda$ of this system we have the equation
\begin{equation}
(\Omega-\lambda)(\Omega-2\Omega Z_0
-\lambda)+\Omega^2 Z_0=0.
\label{5.9}
\end{equation}
Obviously, the characteristic directions (characteristic velocities) are
\begin{equation}
\lambda_1=\Omega (1-Z_0)+\Omega
\sqrt{Z_0^2-Z_0},\quad
   \lambda_2=\Omega (1-Z_0)-\Omega \sqrt{Z_0^2-Z_0}.\quad
\label{5.10}
\end{equation}

We restrict our attention to the case
\begin{equation}
Z_0 > 1.
\label{5.11}
\end{equation}
Then the system (\ref{5.7}), (\ref{5.8}) is a hyperbolic system
and there are two characteristics
\begin{equation}
\eta-\lambda_1 \tau={\rm const},\quad \eta-\lambda_2 \tau={\rm
const}.
\label{5.12}
\end{equation}

We seek a solution of (\ref{5.5}), (\ref{5.6}) in the vicinity of
the characteristic with characteristic velocity $\lambda_1>0$. Using the asymptotic  multiscale expansions method we introduce new variables (the index 1 is omitted)
\begin{equation}
x=\eta-\lambda \tau, \quad T=\mu\tau
\label{5.13}
\end{equation}
and the differential operators
\begin{equation}
\frac{\partial}{\partial\tau}=\mu
\frac{\partial}{\partial T} -\lambda \frac{\partial}{\partial x},  \quad
\frac{\partial}{\partial\eta}=\frac{\partial}{\partial x}.
\label{5.14}
\end{equation}

In general case the functions $h$, $w$ are functions of the old and new variables, i.e. $h=h(\eta,\tau,x,T)$,
$w=w(\eta,\tau,x,T)$. However,  it can be shown that for a closed system of equations it is sufficient that these functions depend only on the new
variables.

We seek the solution in the form
\begin{eqnarray}
   && h=h^0(x,T)+\mu h^1(x,T)+\dots,  \nonumber \\
   && w=w^0(x,T)+\mu w^1(x,T)+\dots \label{5.15}
\end{eqnarray}

For convenience we divide equation (\ref{5.5}) on $c^2(1+\mu
h)$ and omit the terms of order
$O(\mu^2)$
\begin{equation}
 h_\tau + \Omega h_\eta+ \frac{Z_0}{c^2} w_\eta+
\frac{\mu}{c^2}(1-Z_0)h w_\eta + \frac{\mu}{c^2} w h_\eta
=\frac{1}{6}\varepsilon^2 \frac{Z_0^3}{c^2} w_{\eta\eta\eta},
\label{5.16}
\end{equation}
\begin{eqnarray}
&& w_\tau
-2\Omega Z_0 w_\eta -2\Omega \mu (hw)_\eta -\Omega^2 c^2 h_\eta -\Omega^2 c^2 \mu h
h_\eta + \Omega w_\eta +
\frac{\mu}{c^2} w w_\eta
=\nonumber\\
&&=\varepsilon^2\left(
\frac{1}{2}w_{\eta\eta\tau}
+\frac{1}{6}\Omega w_{\eta\eta\eta}
\right).
\label{5.17}
\end{eqnarray}

Substituting (\ref{5.15}) into (\ref{5.16}), (\ref{5.17}) and
using (\ref{5.14}) we obtain
\begin{eqnarray}
&&\left(\mu \frac{\partial}{\partial T} -\lambda
\frac{\partial}{\partial x}\right)
(h^0+\mu h^1)+ \Omega (h^0+\mu h^1)_x+
\frac{Z_0}{c^2}(w^0+\mu w^1)_x+\nonumber\\
&&+\frac{\mu}{c^2}(1-Z_0)(h^0+\mu h^1)(w^0+\mu w^1)_x+
\frac{\mu}{c^2} (w^0+\mu w^1)(h^0+\mu h^1)_x=\nonumber\\
&&=\frac{1}{6}\mu \frac{Z_0^3}{c^2} (w^0+\mu w^1)_{xxx},
\label{5.18}
\end{eqnarray}
\begin{eqnarray}
&&\left(\mu \frac{\partial}{\partial T} -\lambda
\frac{\partial}{\partial x}\right)(w^0+\mu w^1)- \nonumber \\
&&{}-2\Omega Z_0 (w^0+\mu w^1)_x-2\Omega \mu ((h^0+\mu h^1)(w^0+\mu w^1))_x-\nonumber\\
&&-\Omega^2 c^2 (h^0+\mu h^1)_x-
\mu \Omega^2 c^2 (h^0+\mu h^1)(h^0+\mu h^1)_x+
\Omega (w^0+\mu w^1)_x+\nonumber\\
&&+\frac{\mu}{c^2} (w^0+\mu w^1)(w^0+\mu w^1)_x
=\nonumber\\
&&=\mu\left(
\frac{1}{2}
\left(\mu \frac{\partial}{\partial T} -\lambda
\frac{\partial}{\partial x}\right)
(w^0+\mu w^1)_{xx}
+\frac{1}{6}\Omega (w^0+\mu w^1)_{xxx}
\right).
\label{5.19}
\end{eqnarray}

Collecting terms of the same powers of $\mu$, we have
\begin{eqnarray}
&&-\lambda h^0_x +\Omega h^0_x + \frac{Z_0}{c^2} w^0_x=0,
\label{5.20}\\
&&-\lambda w^0_x -2\Omega Z_0 w^0_x + \Omega  w^0_x -\Omega^2 c^2 h^0_x=0.
\nonumber
\end{eqnarray}

\begin{eqnarray}
&&-\lambda h^1_x +\Omega h^1_x + \frac{Z_0}{c^2} w^1_x=
\label{5.21}\\
&&=-
\left\{
h^0_T+\frac{1-Z_0}{c^2} h^0 w^0_x + \frac{1}{c^2} w^0 h^0_x-
\frac{1}{6} \frac{Z_0^3}{c^2} w^0_{xxx}
\right\}\equiv f_1,\nonumber\\
&&-\lambda w^1_x -2\Omega Z_0 w^1_x + \Omega  w^1_x -\Omega^2 c^2 h^1_x=\nonumber\\
&&=-
\left\{
w^0_T-2\Omega (h^0 w^0)_x +\frac{1}{c^2} w^0 w^0_x -\Omega^2 c^2 h^0 h^0_x+
\frac{1}{2} \lambda w^0_{xxx}-\frac{1}{6} \Omega w^0_{xxx}
\right\}\equiv f_2.
\nonumber
\end{eqnarray}

For obtaining the derivatives $h^0_x$, $w^0_x$ we have
the linear system (\ref{5.20}). The matrix of this system is  degenerated and has the form
\begin{equation}
M= \left(
\begin{array}{cc}
 \Omega Z_0-\Omega\sqrt{Z_0^2-Z_0}, & Z_0 c^{-2}\\
  -\Omega^2 c^2,                    & -\Omega Z_0-\Omega\sqrt{Z_0^2-Z_0},
\end{array}
\right).
\label{5.22}
\end{equation}

The left eigenvector ($\ell_1,\ell_2$) of the matrix $M$ or the
eigenvector of the adjoint matrix is given by relations
\begin{eqnarray}
&&  \left(\Omega Z_0-\Omega\sqrt{Z_0^2-Z_0}\right)\ell_1-\Omega^2 c^2\ell_2=0,\nonumber\\
&&\ell_1=\Omega^2 c^2, \quad \ell_2=\Omega Z_0-\Omega\sqrt{Z_0^2-Z_0}.
\label{5.23}
\end{eqnarray}

The solvability condition of  the system (\ref{5.21}) has the form
\begin{equation}
f_1 \ell_1+ f_2 \ell_2=0.
\label{5.24}
\end{equation}

Integrating (\ref{5.21}) with respect to $x$ we obtain
(arbitrary function of time $T$ is omitted)
\begin{equation}
w^0=N h^0, \quad
N=\frac{c^2(\lambda-\Omega)}{Z_0}=-~\frac{c^2 \ell_2}{Z_0}. \label{5.25}
\end{equation}

Using (\ref{5.25}) and the solvability condition (\ref{5.24}) we
have
\begin{eqnarray}
&& \left\{ h^0_T+\frac{1-Z_0}{c^2} N h^0 h^0_x + \frac{1}{c^2}
N h^0 h^0_x- \frac{1}{6} \frac{Z_0^3}{c^2} N h^0_{xxx}
\right\}\ell_1+\label{5.26}\\
&&
+N\left\{ h^0_T-4\Omega h^0 h^0_x +\frac{1}{c^2} N h^0 h^0_x -\Omega^2 c^2 h^0 h^0_x+
\frac{1}{2} \lambda h^0_{xxx}-\frac{1}{6} \Omega h^0_{xxx}
\right\}\ell_2
    =0.
\nonumber
\end{eqnarray}

Finally,
using formulae (\ref{5.23}) for $\ell_1$, $\ell_2$ and (\ref{5.25}) for $N$ we obtain the Korteweg-de Vries equation (index 0 is omitted)
\begin{equation}
h_T+\alpha_0 h h_x + \beta_0 h_{xxx}=0,
\label{5.27}
\end{equation}
where
\begin{eqnarray}
 \alpha_0&=&\frac{\ell_1 (2-Z_0)c^{-2} N -4\Omega N \ell_2 +N^2 c^{-2}
\ell_2-\Omega^2 c^2 \ell_2}
{\ell_1+N \ell_2}=
\label{5.28}\\
&=&\frac{3\Omega(Z_0-\sqrt{Z_0^2-Z_0})(2Z_0-1-2\sqrt{Z_0^2-Z_0})}
{2Z_0(1-Z_0+\sqrt{Z_0^2-Z_0})},
\nonumber
\end{eqnarray}
\begin{eqnarray}
&&\beta_0=\frac{-Z_0^3c^{-2}N\ell_1+3\lambda N \ell_2 -\Omega N \ell_2}
 {6(\ell_1+N \ell_2)}=
 \label{5.29}\\
&&=\frac{-\Omega(Z_0-\sqrt{Z_0^2-Z_0})(-Z_0^3+5Z_0+(6Z_0-2)\sqrt{Z_0^2-Z_0}-6Z_0^2)}
{12Z_0(1-Z_0+\sqrt{Z_0^2-Z_0})}.
\nonumber
\end{eqnarray}

Now we return to the original variables (see, (\ref{4.1}), (\ref{5.3}), (\ref{5.13})) and introduce the notations
\begin{equation}
\Theta=\theta-\lambda t, \quad
\lambda=\Omega(1-Z_0+\sqrt{Z_0^2-Z_0}),\nonumber
\end{equation}
\begin{equation}
H=\mu h, \quad
R=c(1+H). \label{5.30}
\end{equation}

In this case the KdV equation takes the form
\begin{equation}
H_t+\alpha_0 H
H_\Theta + \beta_0 H_{\Theta\Theta\Theta}=0.
\label{5.31}
\end{equation}

This equation describes the perturbation of the free surface in a
coordinate system moving with velocity $\lambda$ and it can be used to characterize the nonlinear behaviors of the film flow traveling waves.

In the fixed coordinate system it is possible to write
\begin{equation}
H_t+\lambda H_\theta+\alpha_0 H H_\theta + \beta_0
H_{\theta\theta\theta}=0.
\label{5.32}
\end{equation}

The  result obtained means that the KdV equation is a rough model of liquid film motion. Note, that we perform surface curvature effect with help constant coefficients $\alpha_0$ and $\beta_0$.

Other words, the previous Boussinesq model describes more accurately the behavior of thin liquid  film on the rotating cylinder
surface. However, the rough model also allows to describe the behavior of the liquid film. In particular, with help of the rough model we can describe the precessing liquid film motion in the azimuthal direction.

\setcounter{equation}{0}

\section{The Painlev\'{e} Analysis}\label{zhmor:sec:6}

In this section we present the algorithm for the well-known Painlev\'{e} integrability test \cite{ConteMusette,Ramani}, which may greatly aid the investigation of integrability and the search for exact solutions. The more general definition used by Painlev\'{e} himself requires all solutions of the ODE to be single-valued around all movable singularities. A later version \cite{BaldwinHereman} allows testing of PDEs directly without reducing them to ODEs.

\subsection{The algorithm and implementation}\label{zhmor:sec:6.1}

We assume a Laurent expansion for the solution $H_{i}$
\begin{equation}
H_{i}(\textbf{x})=\phi^{\alpha_{i}}(\textbf{x})
\sum^{\infty}_{k=0}H_{i,k}(\textbf{x})\phi^{k}(\textbf{x})
, \quad H_{i,0}(\textbf{x})\neq0 \quad \text{and} \quad \alpha_{i}\in Z^{-},
\label{6.1}
\end{equation}
where $H_{i,k}(\textbf{x})$ is an analytic function in the neighborhood of $\phi(\textbf{x})$ .

The solution
should be single-valued in the neighborhood of the non-characteristic, movable
singular manifold $\phi(\textbf{x})$, which can be viewed as the surface of the movable
poles in the complex plane.
The algorithm for the Painlev\'{e} test is composed of the following four steps:

\medskip

\textbf{Step 1:} (determination of the dominant behavior). To determine the strictly negative integer $\alpha_{i}$ and the function $H_{i,0}(\textbf{x})$ we substitute
\begin{equation}
H_{i}(\textbf{x})=H_{i,0}(\textbf{x})\phi^{\alpha_{i}}(\textbf{x}) \label{6.2}
\end{equation}
into the following systems of $M$ polynomial differential equations
\begin{equation}
\Delta(\textbf{H}(\textbf{x}),\textbf{H}'(\textbf{x}),\textbf{H}''(\textbf{x}),\dots,
\textbf{H}^{(m)}(\textbf{x}))=0,
\label{6.3}
\end{equation}
where the dependent variable $\textbf{H}$ has $M$ components $H_{i}$, the independent variable \textbf{x} has $N$ components $x_{j}$,  and $\textbf{H}^{(m)}(\textbf{x})$ denotes the collection of mixed derivative terms of order $m$.

In the resulting polynomial system the equating of every two possible lowest exponents of $\phi(\textbf{x})$ in each equation gives a linear system for determination of $\alpha_{i}$.

If one or more exponents $\alpha_{i}$ remain undetermined then we assign a strictly negative integer value to the free $\alpha_{i}$ so that every equation in (\ref{6.3}) has at least two different terms with equal lowest exponents. Once $\alpha_{i}$ is known, we substitute (\ref{6.2}) into (\ref{6.3}) and solve for $\textbf{H}^{(m)}(\textbf{x})$.

\medskip

\textbf{Step 2:} (determination of the resonances). For each $\alpha_{i}$ and $H_{i,0}(\textbf{x})$ we calculate the integer $r$ for which $H_{i,r}(\textbf{x})$ is an arbitrary function in (\ref{6.1}). We substitute
\begin{equation}
H_{i}(\textbf{x})=H_{i,0}(\textbf{x})\phi^{\alpha_{i}}(\textbf{x})+
H_{i,r}(\textbf{x})\phi^{\alpha_{i}+r}(\textbf{x})
\label{6.4}
\end{equation}
into (\ref{6.3}), keeping only the most singular terms in $\phi(\textbf{x})$, and require the equating of the coefficients  $H_{i,r}(\textbf{x})$ to zero. It is correspond to determination of the roots of $\operatorname{det}Q = 0$, where the $M\times M$ matrix $Q$ satisfies
\begin{equation}
Q.\textbf{H}_{r}=\textbf{0},  \quad  \textbf{H}_{r}=(H_{1,r}, H_{2,r},\dots,H_{M,r})^{T}. \label{6.5}
\end{equation}

\medskip
\textbf{Step 3:} (determination of the integration constants and checking of the compatibility
conditions). To possess the Painlev\'{e} property the arbitrariness of $H_{i,r}(\textbf{x})$ must be verified up to the highest resonance level, i.\,e. all compatibility conditions must be trivially satisfied.

To verify these conditions we substitute
\begin{equation}
H_{i}(\textbf{x})=\phi^{\alpha_{i}}(\textbf{x})
\sum^{r_{Max}}_{k=0}H_{i,k}(\textbf{x})\phi^{k}(\textbf{x}). \label{6.6}
\end{equation}
into (\ref{6.3}), where $r_{Max}$ is the highest positive integer resonance.

For the system with the Painlev\'{e} property the quantity of the arbitrary constants of integration at resonance levels must be coincide with the quantity of resonances at that level. Furthermore, all constants of integration $H_{i,r}(\textbf{x})$ at non-resonance levels must be clearly determined.


\medskip
\textbf{Step 4:} (generation of the truncated expansion). For each pair of $(\alpha_{i},H_{0}(\textbf{x}))$ we
calculate the possible truncated expansion in the form
\begin{equation}
H(\textbf{x})=H_{0}\phi^{\alpha_{i}}+H_{1}\phi^{\alpha_{i}+1}+
\dots +H_{-\alpha_{i}}\phi^{0}, \quad  i=1,\dots,M,
\label{6.7}
\end{equation}
where $H_{l}, (l=0,\dots,-\alpha_{i}-1)$ can be determined with  substituting (\ref{6.7}) into (\ref{6.3}) and equating coefficients of the identical power of $\phi$.

If $H_{l}, (l=0,\dots,-\alpha_{i}-1)$ cannot be determined, then the series (\ref{6.1}) cannot be truncated at constant
terms.


\subsection{The Painlev\'{e} test of the KdV equation}\label{zhmor:sec:6.2}

To determine the dominant behavior, we substitute (\ref{6.2}) into the KdV equation and remove the exponents of $\phi(\theta,t)$. The removing of duplicates and non-dominant exponents, and considering all possible balances of two or more exponents leads to
\begin{equation}
\alpha_{1}=-2.
\label{6.8}
\end{equation}

Substituting $H(\theta, t)=H_{0}(\theta, t)\phi^{-2}(\theta, t)$ into (\ref{5.32}) and solving for $H_{0}(\theta, t)$ we get
\begin{equation}
H_{0}(\theta, t)=-12 \beta \phi_{\theta}^{2}(\theta, t). \label{6.9}
\end{equation}

Substituting $H(\theta, t)=-12 \beta \phi_{\theta}^{2}(\theta, t) \phi^{-2}(\theta, t)H_{r}(\theta, t)\phi^{r-2}(\theta, t)$ into (\ref{5.32}), keeping the most singular terms, and taking the coefficient of $H_{r}(\theta, t)$ we get
\begin{equation}
r=-1, 4  \quad \text{and}  \quad 6.
\label{6.10}
\end{equation}

While we are only concentrated on the positive resonances. The value $r=-1$ is so called the universal resonance and corresponds to the arbitrariness of the manifold $H(\theta, t)$.
The constants of integration at level $j$ are found with  help of the substituting (\ref{6.6}) into (\ref{5.32}), where $r_{Max}=6$,  and by the removing of the coefficients at $\phi^{j}(\theta, t)$, such $H_{4}(\theta, t)$  and $H_{6}(\theta, t)$ are arbitrary functions of $t$ since the conditions at resonance $r=4$, and $r=6$ are satisfied.

To construct the B\"{a}cklund transformation of equation (\ref{5.32}) we truncate the Laurent series according to Step 4 at the constant level term
\begin{equation}
H(\theta, t)=H_{0}(\theta, t)\phi^{-2}(\theta, t)+H_{1}(\theta, t)\phi^{-1}(\theta, t)+H_{2}(\theta, t). \label{6.11}
\end{equation}

Therefore, we obtain an auto-B\"{a}cklund transformation of equation (\ref{5.32}) as follows
\begin{equation}
H=\frac{-12 \beta \phi_{\theta}^{2}(\theta, t)}{\phi^{2}(\theta, t)}+\frac{12 \beta \phi_{\theta\theta}(\theta, t)}{\phi(\theta, t)}+H_{2},
\label{6.12}
\end{equation}
where $H_{2}$ is a solution of the KdV equation.

We take the vacuum solution at $H_{2}=0$ in equation (\ref{6.11}) which leads to
\begin{equation}
H=H_{0}\phi^{-2}+H_{1}\phi^{-1}=12\beta\,\frac{\partial^{2}}{\partial x^{2}}\,(\ln\phi). \label{6.13}
\end{equation}
Using the above auto-B\"{a}cklund transformation and choosing the different $H(\theta, t)$ and $\phi(\theta, t)$, one can obtain various solutions (as in \cite{BaldwinHereman}).
We can derive the special solution  inserting (\ref{6.13}) into (\ref{5.32}) and collecting the terms with the identical power of $\phi$. We get a system of homogeneous PDEs for $\phi$.

Finally, we get the following solution
\begin{equation*}
\phi=c_2
\left(
\frac{3}{8}
(\xi-4 \beta c_1)+
\frac{1}{\gamma} \sin
\frac{\gamma(\xi -4 \beta c_1)}{2}
+
\frac{1}{8\gamma}
\sin
\gamma(\xi -4 \beta c_1)
\right)+c_3,
\end{equation*}
\begin{equation}
\gamma=\frac{\sqrt{-V+c_0}}{\sqrt{\beta}},
\label{6.14}
\end{equation}
where $c_1,c_2$ and $c_3$ are arbitrary constants.

\begin{figure}[H]
\centering  \includegraphics[scale=0.6]{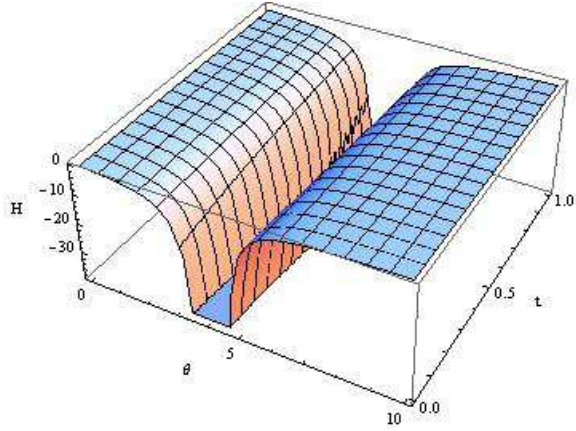}\\
  \caption{Solitary wave solution of the KdV model equation by using the auto-B\"{a}cklund transformation}\label{fig1}
\end{figure}

This result means
that the solution of the thin liquid film model corresponded to the B\"{a}cklund transformation that is a solitary wave solution as shown in Fig.~\ref{fig1}.

\begin{figure}[H]
\centering
\includegraphics[scale=0.6]{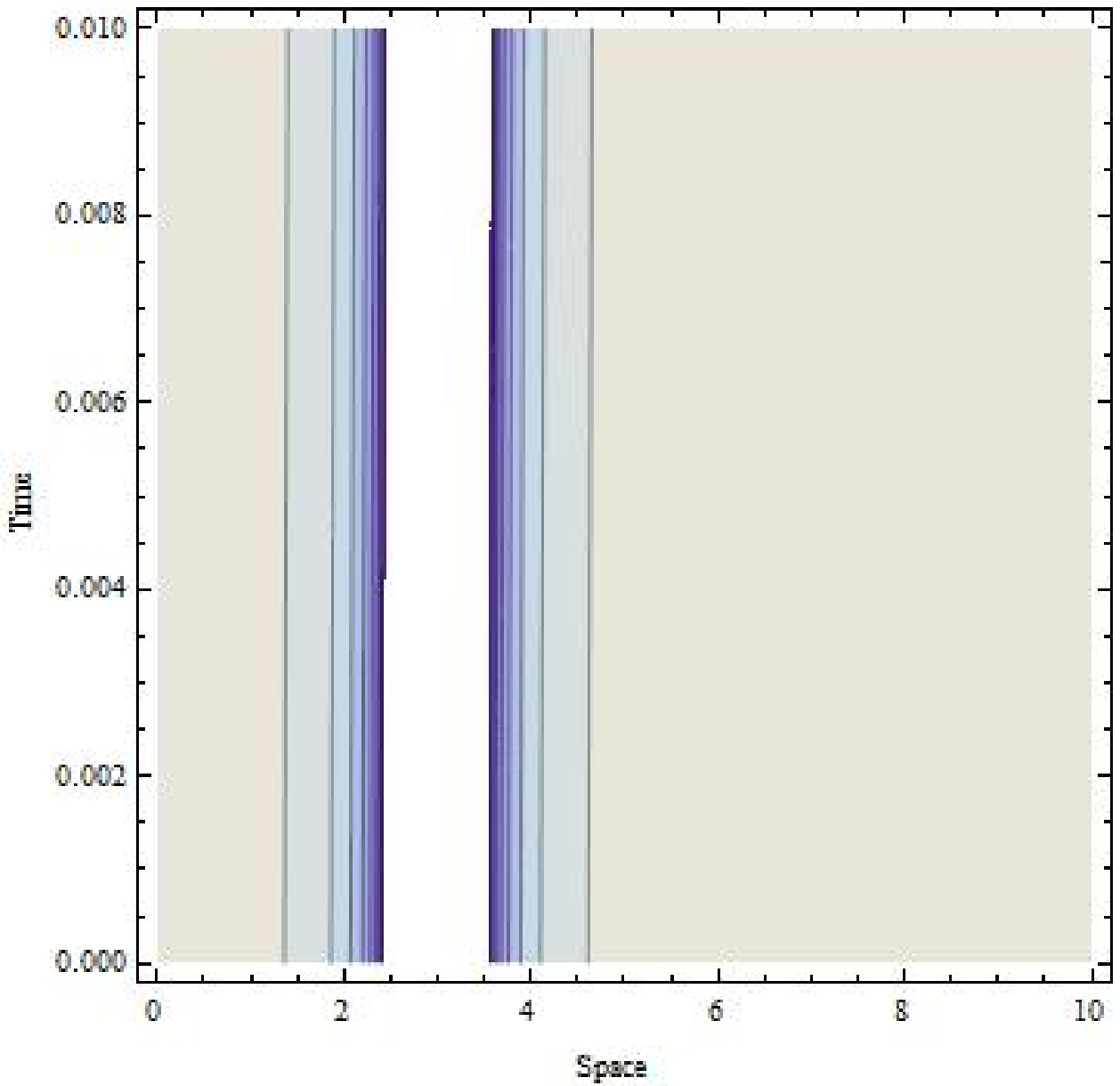}\
\includegraphics[width=3 in]{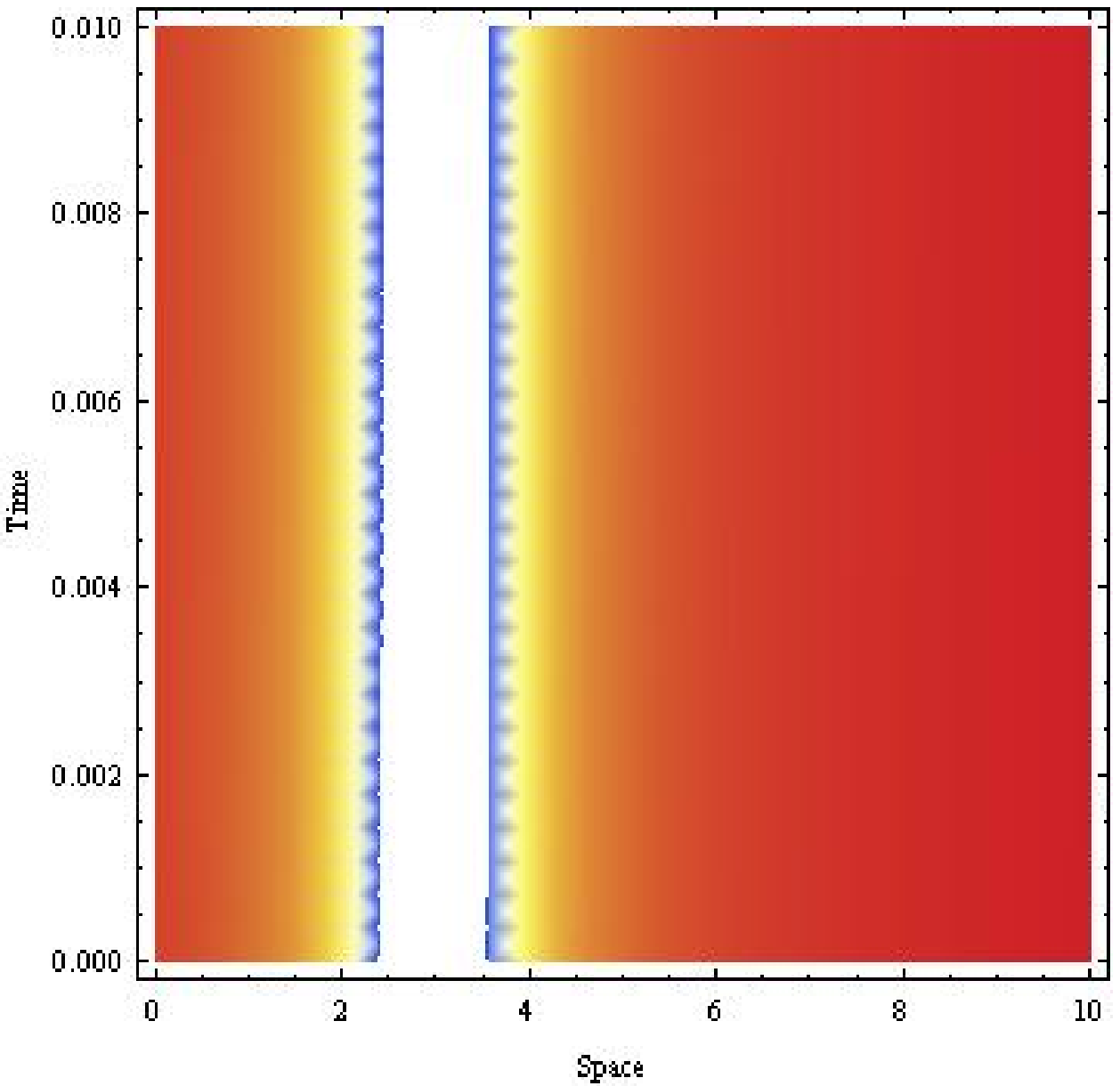}
  \caption{The perturbation of the free surface of a rotating cylinder by B\"{a}cklund transformation}\label{fig2}
\end{figure}

Fig.~\ref{fig2} shows the perturbation of the free surface of a rotating cylinder which describes the behavior of the thin liquid film.

\setcounter{equation}{0}

\section{Analytical solution of the KdV equation}\label{zhmor:sec:7}

To determine the solutions for various cases studied in this paper analytical schemes are used.  The results for the nonlinear PDE (\ref{5.32}) are presented and compared with the previous works and experimental situation \cite{AbourabiaDanafMorad,ChenChenYang,SirwahZakaria}.

\subsection{Periodic solution of the KdV equation}\label{zhmor:sec:7.1}

\medskip
We introduce a self-similar variable $\xi$ and the following
notations
\begin{equation}
\xi=\theta-\Omega_0 t, \quad \Omega_0=\alpha_0 V,
\quad \lambda=c_0 \alpha_0, \quad \beta_0=\alpha_0 \beta.
\label{7.1}
\end{equation}
In this case equation (\ref{5.32}) has the following form
\begin{equation}
-V H'+c_0 H'+H H'+ \beta H'''=0.
\label{7.2}
\end{equation}
Integrating (\ref{7.2}) twice we get
\begin{equation}
3\beta
(H')^2=(b_1-H)(b_2-H)(b_3-H).
\label{7.3}
\end{equation}
Here $b_k$ are the
constants of integration connected by relation
\begin{equation}
V-c_0=\frac{b_1+b_2+b_3}{3}.
\label{7.4}
\end{equation}

The periodic solution of (\ref{7.3}) takes the form
\begin{equation}
H(\xi)=\frac{2b}{s^2}\operatorname{dn}^2(z;s)+b_3,
\label{7.5}
\end{equation}
where
\begin{equation}
b=\frac{b_1-b_2}{2}, \quad
s^2=\frac{b_1-b_2}{b_1-b_3}, \quad
z=\left(\frac{b}{6\beta}\right)^{1/2}\frac{\xi}{s}.
\label{7.6}
\end{equation}

Here $\operatorname{dn}^2(z;s)$ is the Jacobi elliptic function
\begin{eqnarray}
&&
\operatorname{dn}^2(z;s)=\frac{\pi}{2K(s)}+
\frac{2\pi}{K(s)}\sum\limits_{n=1}^{\infty}\frac{q^n}{1+q^{2n}}cos\frac{\pi n z}{K(s)},\nonumber\\
&& q=\exp\left\{-\pi\frac{K(s')}{K(s)}\right\}, \quad s'=\sqrt{1-s^2},
\label{7.7}
\end{eqnarray}
where $K(s)$ is the complete elliptic integral of the first kind.

The function $\operatorname{dn}^2(z;s)$ has a period $2K(s)$ and the wavelength $\Lambda$ of
the periodic solution is given by
\begin{equation}
\Lambda=2\left(\frac{6\beta}{b}\right)^{1/2}s K(s).
\label{7.8}
\end{equation}

In the case of $m$ waves on a circle we have
\begin{equation}
\Lambda=\frac{2\pi}{m}.
\label{7.9}
\end{equation}

The precession velocity $V$ of the
periodic solution and the average value $\bar H$ are given by
\begin{equation}
V=c_0+\frac{2b(2-s^2)}{3s^2}+b_3,
\label{7.10}
\end{equation}
\begin{equation}
\bar H=\Lambda^{-1}\int\limits_{0}^{\Lambda}H(\xi)d\,\xi=
\frac{2b}{s^2}\frac{E(s)}{K(s)}+b_3,
\label{7.11}
\end{equation}
where $E(s)$ is the complete elliptic integral of the second kind.

The minimum and maximum values of the function $H(\xi)$ are defined by the formulae
\begin{equation}
H_{min}=b_2, \quad H_{max}=b_1.
\label{7.12}
\end{equation}

\begin{figure}[H]
\centering  \includegraphics[scale=0.7]{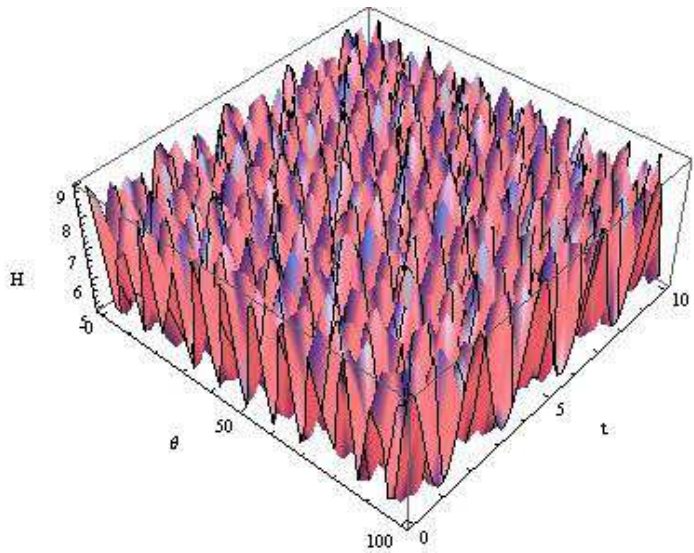}\\
\caption{The periodic solitary wave solution of the KdV equation} \label{fig6}
\end{figure}

\begin{figure}[H]
\centering  \includegraphics[scale=1.3]{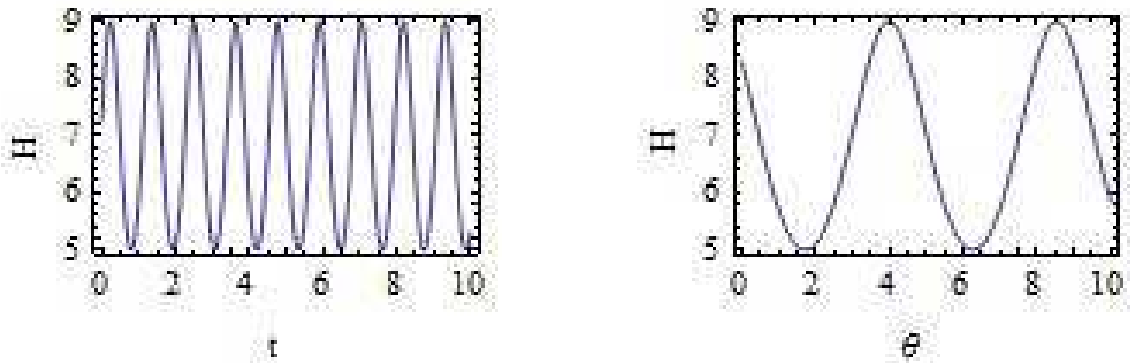}\\
\caption{The perturbation of the free surface of a rotating cylinder} \label{fig7}
\end{figure}

The periodic solitary wave solution (\ref{7.5}) is shown
on Figs.~\ref{fig6}, \ref{fig7}.

\subsection{The Tanh-function method}\label{zhmor:sec:7.2}

We use the transformation
\begin{equation*}
H(\theta, t)=H(\xi),
\end{equation*}
where the wave variable
\begin{equation*}
\xi=\theta-\Omega_0 t.
\end{equation*}
This transformation converts the nonlinear PDE (\ref{5.32}) to the equivalent ODE (\ref{7.2}).

The main idea of the Tanh-function method is to use a new variable tanh that allows to represent all derivatives of a tanh by a tanh itself.

Introducing a new independent variable (and parameter $\nu$)
\begin{equation}
Y=\tanh[\nu\xi]
\label{7.13}
\end{equation}
leads to the change of derivatives:
\begin{eqnarray}
&& \frac{d}{d\xi}=\nu(1-Y^2)\frac{d}{d Y},\label{7.15}\\
&&\frac{d^2}{d\xi^2}=
\nu^2Y(1-Y^2)
\left(
-2Y\frac{d}{d Y}+\frac{d^2}{d Y^2}
\right),
\nonumber\\
&&\frac{d^3}{d\xi^3}=
\nu^3Y(1-Y^2)
\left(
(6Y^2-2)\frac{d}{d Y}-6Y(1-Y^2)\frac{d^2}{d Y^2}+(1-Y^2)^2\frac{d^3}{d Y^3}
\right).
\nonumber
\end{eqnarray}

We construct solutions with help of the following finite series expansion
\begin{equation}
H(\xi)=S(Y)=\sum^{m}_{i=0}a_{i}Y^{i}
\label{7.14}
\end{equation}
in which all real constants $a_{i}$ should be determined later.

The positive integer parameter $m$ is  obtained
by balancing the linear terms of the highest order in the equation with the highest order nonlinear terms \cite{Yan}.
The highest degree of the linear term $d^jH/d\xi^j$ is taken as
\begin{equation}
\operatorname{deg}\left(\frac{d^jH}{d\xi^j}\right)=m+j,\quad j=1,2,3,\dots \label{7.16}
\end{equation}
and the nonlinear term $H^k d^jH/d\xi^j$ is taken as
\begin{equation}
\operatorname{deg}\left(H^k \frac{d^jH}{d\xi^j}\right)=(k+1)m+j,\quad k=0,1,2,3,\dots
\label{7.17}
\end{equation}

\textit{\textbf{We give a brief description of the tanh method as follows:}}

\medskip

\textbf{Step 1} We determine the parameter $m$ by balancing the highest-order partial derivative term and the nonlinear term in equation (\ref{5.32}).

\textbf{Step 2} Using the standard mathematical (symbolic) or numerical software we substitute (\ref{7.14}) into equation (\ref{7.2}), equating to zero the coefficients of all power $Y^i$ yields an over-determined system of nonlinear algebraic equation for $a_{i}, \nu$.

\medskip

\begin{figure}[H]
\centering
 \includegraphics[scale=0.6]{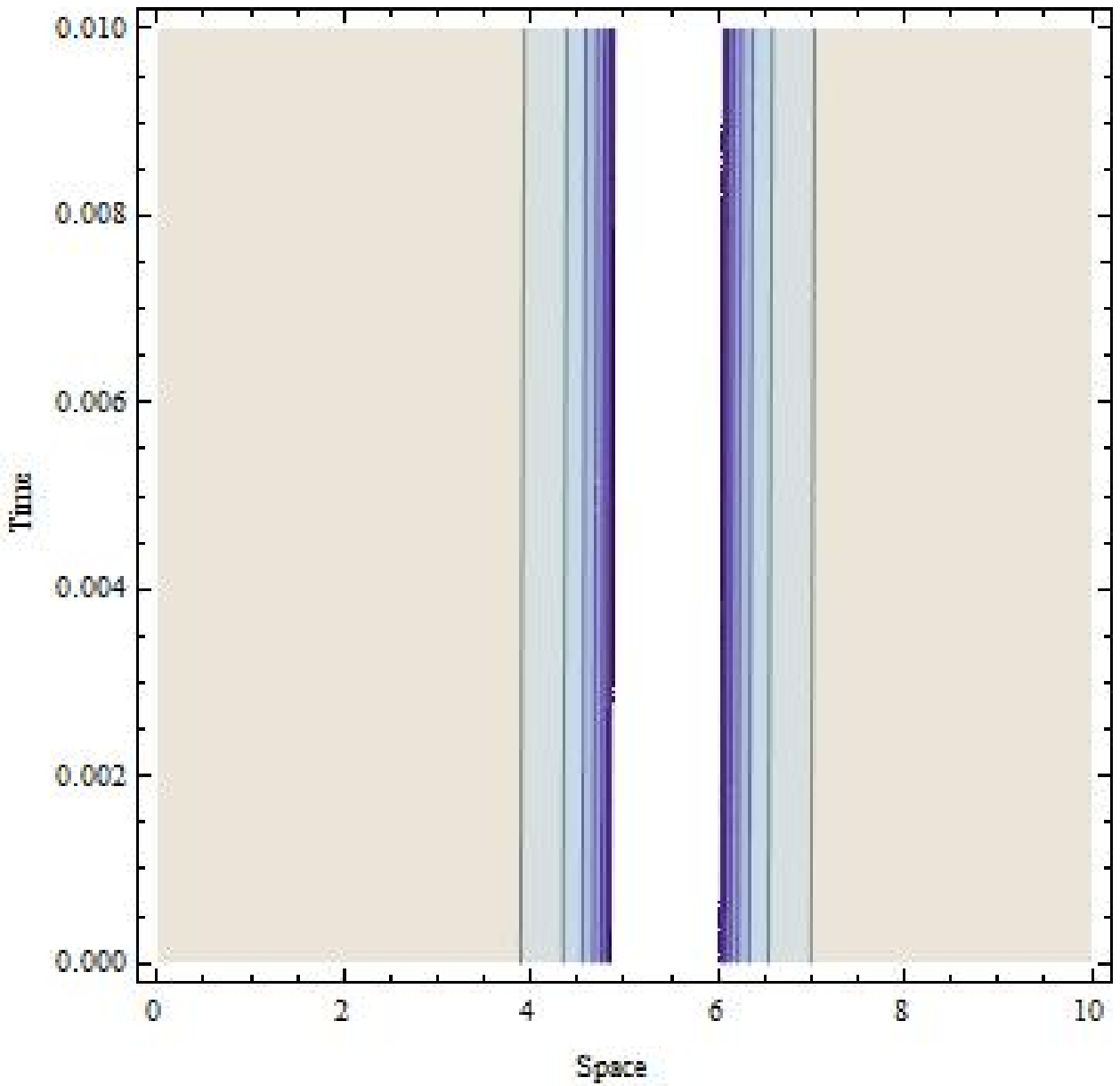}\
 \includegraphics[scale=0.6]{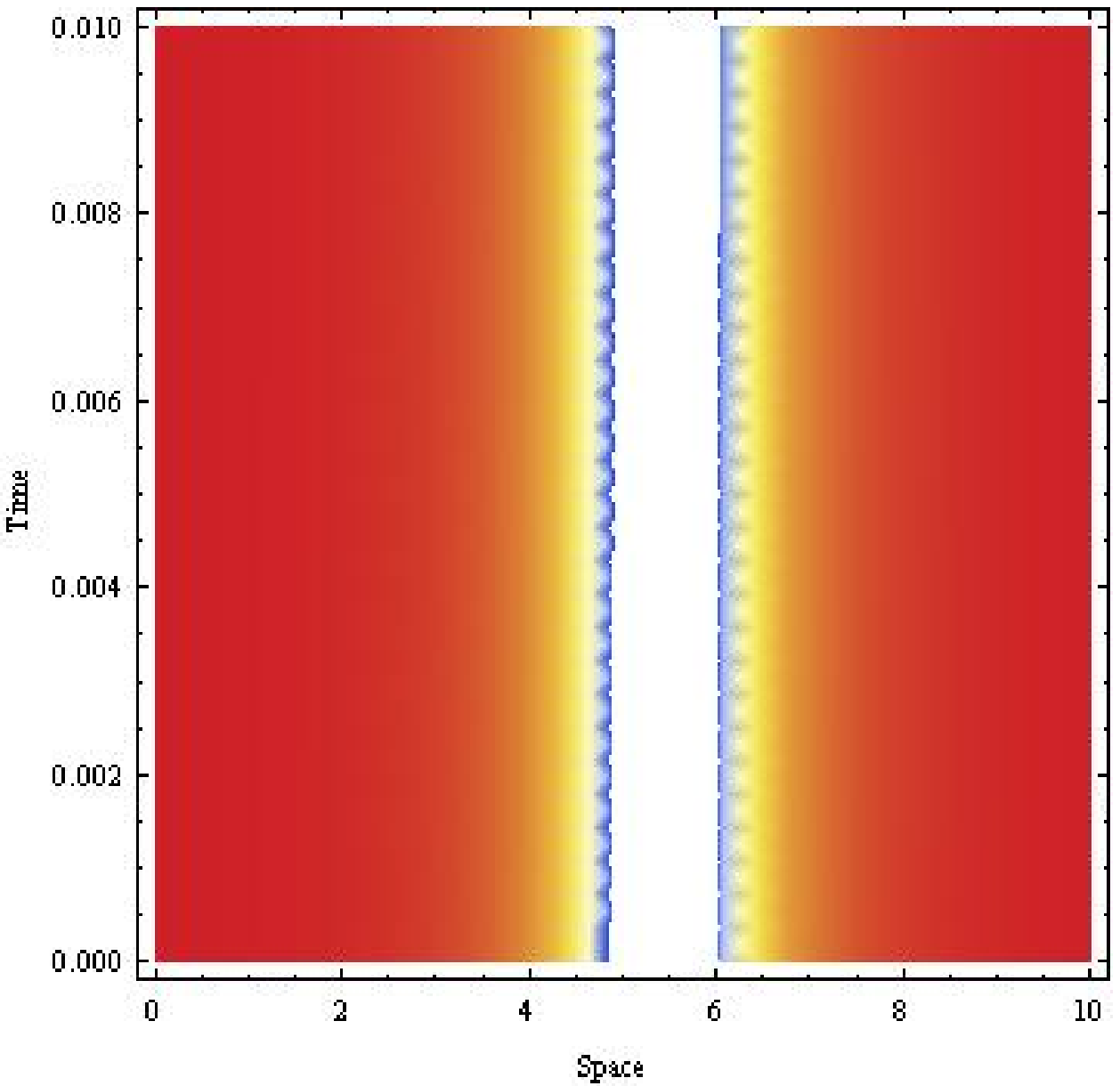}\
  \caption{The perturbation of the free surface of a rotating cylinder}\label{fig3}
\end{figure}

\begin{figure}[H]
\centering
 \includegraphics[scale=0.6]{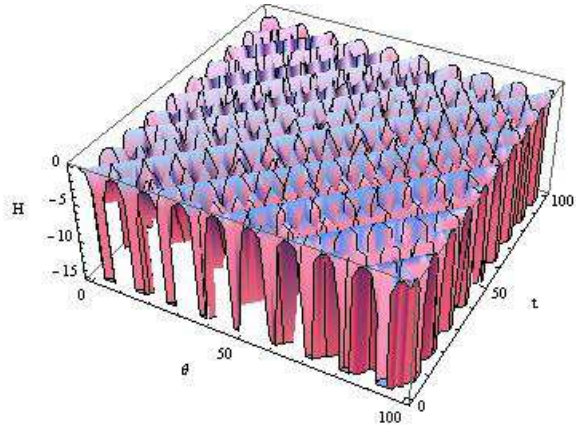}\\
  \caption{Solitary wave solution of the KdV model equation by using the Tanh-function method}\label{fig4}
\end{figure}

We consruct the solution of the 1D KdV equation (\ref{5.31}) by using the abovementioned method. We can deduce from (\ref{7.16}) and (\ref{7.17}) into (\ref{5.32}) that $m=2$. It means that the KdV equation (\ref{5.32}) may have the following traveling wave solution
\begin{equation}
H(\xi)=a_{0}+a_{1}Y+a_{2}Y^{2}.
 \label{7.18}
\end{equation}
Substituting (\ref{7.18}) into (\ref{7.2}), yields an algebraic system for $a_{i}(i=0,1,2)$. The solutions to this algebraic equations can be derived
\begin{equation}
a_0=V-c_0+8\beta\mu^2, \quad a_1=0, \quad a_2=-12\beta\mu^2.
\end{equation}
These solutions represent solitary waves, such that the wave profile is periodic, which gives good agreement with the the Painlev\'{e} analysis and it's B\"{a}cklund transformation as shown in Figs.~\ref{fig3}, \ref{fig4}.

\begin{figure}[H]
\centering
 \includegraphics[scale=0.6]{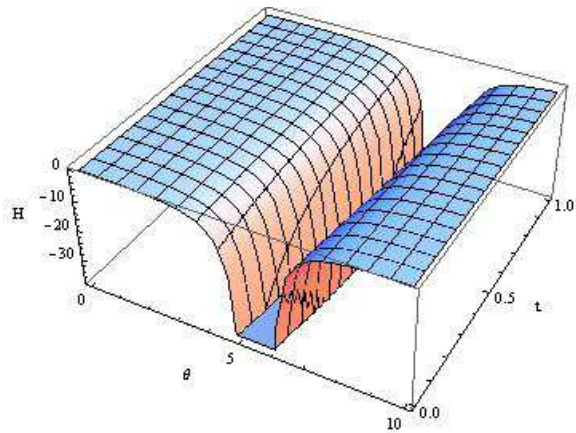}\\
  \caption{The perturbation of the free surface of a rotating cylinder}\label{fig5}
\end{figure}

On the Fig.~\ref{fig3}--\ref{fig5} a rich dynamical behavior characterized by appearance of nonlinear waves is illustrated.
The motion of the film in two spatial dimensions with a rotating cylinder and without surface tension exhibits a solitary waves that corresponds to the decomposition of the nonlinear waves in previous models \cite{AbourabiaDanafMorad,ChenChenYang,SirwahZakaria}.

\section*{Conclusion}

\medskip
In this work we have obtained three models, describing the behavior of a liquid film of an incompressible ideal fluid on the surface of a rotating cylinder. Boussinesq model (\ref{4.18}), (\ref{4.19}), the most accurate one, takes into
account the mean curvature of the liquid layer. The equation of KdV (\ref{5.32}) is a rough model, which allows producing records due to a curvature of constant coefficients. The system of quasilinear
equations (\ref{4.18}), (\ref{4.19}) in the case of hyperbolicity also allows to obtain information about the behavior of the liquid film. Despite the fact that the model (\ref{5.32}) is rough, it nevertheless allows us
to construct a solution corresponding to the precession thin film
along the azimuthal direction, determine the shape of the free
surface (in the form of cnoidal and solitary waves) and the rate of precession.
We have applied directly the PDE Painlev\'{e} test to the KdV equation. An auto-B\"{a}cklund transformation is presented by using of the truncated Painlev\'{e} expansion and symbolic computation. The Tanh-function, Jacobi elliptic function expansion methods are used to solve the KdV model. The solutions describe the perturbation of free surface. At the onset of this phase transition, the system reveals rich dynamic behavior characterized by the appearance of solitary waves. In addition, we find that the solution which is obtained from the Painlev\'{e} analysis behaves similar to the solution of the KdV equation with the Tanh-function method.

\setcounter{equation}{0}
\renewcommand{\theequation}{A.\arabic{equation}}

\bigskip

{\bf Appendix 1. The derivation of (\ref{3.5})}

\medskip
To determine the function $\psi(x,z)$ we have the following problem
\begin{equation}
\varepsilon^2\,\psi_{xx}+\psi_{zz}=\Omega\,\varepsilon^2, \quad \psi=\psi(x,z,t),
\label{A.1}
\end{equation}
\begin{equation}
\psi(x,0,t)=0,
\label{A.2}
\end{equation}
where $\Omega$ is the constant.

We seek a solution in the form

\begin{equation}
\psi(x,z,t)=
\sum\limits_{k=0}^{\infty}
\varepsilon^{2k}\,\psi_k(x,z,t) + \frac{1}{2}\,\Omega\,\varepsilon^2\,z^2,
\label{A.3}
\end{equation}
\begin{equation}
\psi_k(x,0,t)=0,\,\,\, k=0,1,2,\dots
\label{A.4}
\end{equation}

Substituting  (\ref{A.3}) in (\ref{A.1}) and collecting terms with identical powers of $\varepsilon$ we obtain
\begin{equation}
\psi_{0,zz}(x,z,t)=0,
\label{A.5}
\end{equation}
\begin{equation}
\psi_{k,zz}(x,z,t)=-\psi_{k-1,xx}(x,z,t),\quad k=1,2,\dots
\label{A.6}
\end{equation}

Taking into account the boundary conditions (\ref{A.4}) we obtain
\begin{equation*}
\psi_0(x,z,t)=c_0(x,t)\,z,
\end{equation*}
\begin{equation}
\!\!\!\!\! \psi_k(x,z,t)=\sum\limits_{i=1}^{k+1}
(-1)^{i+1}\,\frac{z^{2i-1}}{(2i-1)!}\,
\frac{\partial^{2i-2}}{\partial x^{2i-2}}
\,c_{k+1-i}(x,t),\,\,k=1,2,\dots,
\label{A.7}
\end{equation}
where $c_i(x,t)$ are arbitrary functions.

Then function $\psi$ has the form
\begin{equation*}
\psi(x,z,t)=\sum\limits_{k=0}^{\infty}\,\varepsilon^{2k}\,
\sum\limits_{i=1}^{k+1}
(-1)^{i+1}\,\frac{z^{2i-1}}{(2i-1)!}\,
\frac{\partial^{2i-2}}{\partial x^{2i-2}}
\,c_{k+1-i}(x,t)+
 \frac{1}{2}\,\Omega\,\varepsilon^2\,z^2.
\end{equation*}

By changing the order of summation ($\sum\limits_{k=0}^{\infty} \sum\limits_{i=1}^{k+1} =
\sum\limits_{i=1}^{\infty} \sum\limits_{k=i-1}^{\infty}$) we obtain

\begin{equation*}
\psi(x,z,t)=\sum\limits_{j=0}^{\infty}\,(-1)^{j}\,\varepsilon^{\,2j}
\,\frac{z^{2j+1}}{(2j+1)!}\,
\frac{\partial^{2j}}{\partial x^{2j}}
\sum\limits_{k=j}^{\infty}\,\varepsilon^{2(k-j)}c_{k-j}(x,t)+
 \frac{1}{2}\,\Omega\,\varepsilon^2\,z^2.
\end{equation*}

Finally, we have
\begin{equation}
\psi(x,z,t)=\sum\limits_{j=0}^{\infty}\,(-1)^{j}\,\varepsilon^{\,2j}
\,\frac{z^{2j+1}}{(2j+1)!}\,
\frac{\partial^{2j}}{\partial x^{2j}}
\,H(x,t)
+\frac{1}{2}\,\Omega\,\varepsilon^2\,z^2,
\label{A.10}
\end{equation}
where
\begin{equation}
H(x,t)=\sum\limits_{i=0}^{\infty}\,\varepsilon^{2i}c_{i}(x,t).
\label{A.11}
\end{equation}

Of course, the formula (\ref{A.10}) is formal. At least, this formula is valid at $\varepsilon=1$. It is easy to verify with help of direct substitution of (\ref{A.10}) into (\ref{A.1}), (\ref{A.2}).

By analogy,  we can obtain a similar formulae for the polar coordinate system.

Let we have the following problem

\begin{equation}
\frac{1}{r} \frac{\partial}{\partial r} r
\frac{\partial \psi}{\partial r}
 + \varepsilon^2\,
\frac{1}{r^2}
\frac{\partial^2 \psi}{\partial \theta^2}
 =0,\quad, \psi=\psi(r,\theta,t),
\label{A.12}
\end{equation}
\begin{equation}
\psi(a,\theta,t)=0.
\label{A.13}
\end{equation}

We introduce the changing of the  variables
\begin{equation}
\xi=\ln \frac{r}{a},\quad
r\frac{\partial}{\partial r}=
\frac{\partial}{\partial \xi}.
\label{A.14}
\end{equation}

In this case the formulae  (\ref{A.12}), (\ref{A.13}) take the form which coincides with  (\ref{A.1}), (\ref{A.2})
\begin{equation}
\psi_{\xi \xi} + \varepsilon^2\,\psi_{\theta \theta} =0,\quad \psi=\psi(\xi,\theta,t),
\label{A.15}
\end{equation}
\begin{equation}
\psi(0,\theta,t)=0.
\label{A.16}
\end{equation}

Obviously, the solution of the problem (\ref{A.12}), (\ref{A.13}) is \begin{equation}
\psi(r,\theta,t)=\sum\limits_{j=0}^{\infty}\,(-1)^{j}\,\varepsilon^{\,2j}
\,\frac{\left(\ln \left(\frac{r}{a}\right) \right)^{2j+1}}{(2j+1)!}
\,
\frac{\partial^{2j}}{\partial \theta^{2j}}
\,H(\theta,t).
\label{A.17}
\end{equation}

\setcounter{equation}{0}
\renewcommand{\theequation}{B.\arabic{equation}}

\bigskip

{\bf Appendix 2. Jacobi elliptic functions}

\medskip

We present a set of formulae used in Sec.~\ref{zhmor:sec:7.1}.
\begin{equation}
\operatorname{dn}^2(\mu t)=1-s^2~\operatorname{sn}^2(\mu t),
\label{B.1}
\end{equation}
\begin{equation}
\frac{d}{dt}\operatorname{dn}^2(\mu t)=-2s^2
\operatorname{sn}(\mu t) \mu \frac{d}{d\mu t} \operatorname{sn}^2(\mu t),
\label{B.2}
\end{equation}
\begin{equation}
\mu t=\int\limits_{0}^{\operatorname{sn}(\mu t)}\frac{dz}{\sqrt{(1-z^2)(1-s^2 z^2)}},
\label{B.3}
\end{equation}
\begin{equation}
\frac{d}{d\mu t} \operatorname{sn}(\mu t)=\sqrt{(1-\operatorname{sn}^2(\mu t))(1-s^2 \operatorname{sn}^2(\mu t))},
\label{B.4}
\end{equation}
\begin{equation}
\left\{\frac{d}{dt} \operatorname{dn}^2(\mu t)\right\}^2=
4 \mu^2(1-\operatorname{dn}^2(\mu t))\operatorname{dn}^2(\mu t)(s^2-1+ \operatorname{dn}^2(\mu t))
\label{B.5}
\end{equation}

Let
\begin{equation}
u=a \operatorname{dn}^2(\mu t)+c.
\label{B.6}
\end{equation}
Then $u$ satisfies the equation
\begin{equation}
\frac{a}{4 \mu^2} (u')^2=(a+c-u)(c-u)(a+c-a s^2-u).
\label{B.7}
\end{equation}

We introduce notations
\begin{equation}
a+c=b_1, \quad a+c-a s^2=b_2, \quad c=b_3, \quad 3\beta=\frac{a}{4 \mu^2}.
\label{B.8}
\end{equation}

In this case equation (\ref{B.7}) becomes
\begin{equation}
3\beta (u')^2=(b_1-u)(b_2-u)(b_3-u).
\label{B.9}
\end{equation}

Also, we have the relations
\begin{equation}
\frac{2b}{s^2}=b_1-b_3, \quad b=\frac{b_1-b_2}{2}, \quad s^2=\frac{b_1-b_2}{b_1-b_3}.
\label{B.10}
\end{equation}

\end{document}